# New compound control algorithm in sliding mode control to reduce the chattering phenomenon: experimental validation

**Mehran Rahmani**

Department of Mechanical Engineering, University of Wisconsin-Milwaukee, WI, USA

E-mail: mrahmani@uwm.edu

* Corresponding author

**Asif Al Zubayer Swapnil**

Department of Mechanical Engineering, University of Wisconsin-Milwaukee, WI, USA

E-mail: aswapnil@uwm.edu

**Abstract:** In this work, a new SMS is proposed to achieve high tracking and suitable robustness. However, the chattering phenomenon should be regarded as the main drawback of the SMC. Therefore, a new compound control algorithm is used for reducing the chattering phenomenon. The applied compound control law constantly evaluates the error and send the correct value to the system. This significantly will reduce the chattering phenomenon—the performance of the control methods validated by applying on a robot arm experimentally.

## 1. Introduction

SMC is a strong controller for robustness and tracking. It's well-known due to robustness when some external noises applied to the system [1-6]. It can effectively suppress them. The novelty of the SMC is based on how suitably choosing SMS. According to this process, many kinds of research have been released to design a suitable SMC for different systems. Xiong et al. [7] introduced by using quantization process, described distributed SMC. The control algorithm is used with uncertainties, inner coupling, and some quantized structures. To use digital communication, a quantizer is produced on the sensor system. Then, an adaptive law is used to prevent the discontinuity of the SMC. For each system, an integral SMS is used on the basis of the filtered signal. To obtain exponential reaching to SMS in a finite time, a set of dynamic

SMC is processed. Consequently, the performance of the produced distributed SMC approach is demonstrated by simulations validations. Herrera et al. [8] used the Alpeter method and SMC to produce a dynamic SMC. A comparison of the suggested method and SMC illustrated the advantages of the proposed structure. The chattering reduced by using the proposed controller. Yu et al. [9] proposed a new control scheme for the piezoelectric actuator to obtain suitable tracking. A particle swarm algorithm is used for the identification of nonlinear model parameters. To position, the proposed structure, SMC, and feedforward methods are applied by using the Bouc-Wen inverse algorithm. Wang et al. [10] introduced incremental nonsingular SMC for nonlinear systems regarding sudden actuator fault, external perturbations, and model uncertainties. This scheme does not include singularity due to free from any negative fractional order. Therefore, simulations illustrate that the proposed scheme is robust against actuator faults in comparison with nonsingular terminal SMC. Li et al. [11] for a singular markovian jump system considered the problem of asynchronous output feedback SMC design. A novel asynchronous dynamic output feedback SMC method is introduced. Zheng et al. [12] to control the robot with perturbations introduced a fuzzy SMC approach. By applying perturbations, the complex dynamic model of the robot is introduced. Then, the new fuzzy SMC is proposed according to SMC and fuzzy control combination. Furthermore, a deep learning method is applied to achieve a precise dynamic model experimentally. It estimates the dynamic model perfectly. Finally, the KUKA robot is selected to verify the performance of the intelligent fuzzy SMC approach, which suitable results obtained by experiments. Jing et al. [13] introduced a novel adaptive SMC to perturbation rejection applied to the robotic manipulator. To guarantee the steady-state and the transient performance for robotic arms, modification based on tracking error by applying certain functions implemented. A nonsingular SMS is implemented by applying the modified error. Then, to stabilize the system, a terminal SMC is used. A new sliding mode observer is applied to suppress external disturbances and compensate for the uncertainties. An adaptive algorithm generated from equivalent control is proposed for considering lumped disturbance. The adaptive SMC is designed in a combination of nonsingular terminal SMC, adaptive algorithm, and sliding mode disturbance observer. Validation of the designed controller is performed by different simulations. Ferrara et al. [14] prepared a controller for an industrial robot manipulator by using a switching method. Two cases are proposed in this controller: inverse dynamic and decentralized methods. The first one is suitable for improving the velocity and acceleration performance, and the second one is convenient for suitable transmission ratios. Therefore, the integral SMC is applied to approximate the unmodeled dynamic and compensate matched perturbations. Yi and Zhai [15] considered a controller when inertia uncertainties and external perturbations are applied to a robotic manipulator. The adaptivity of the SMC makes chattering free for the system. A second-order fast nonsingular terminal SMC is applied to obtain desirable tracking, fast convergence, and to ensure robustness and system performance. It's not required to use the upper bound by applying the adaptive algorithm. Xia et al. [16] to control of uncertain systems with time delay, proposed robust SMC. The robust reaching control algorithm is used for SMS based on the linear matrix. Zhang and Yan [17] to control of piezoelectric, proposed an adaptive observer integral SMC. An adaptive observer is designed to suppress the noises by using a Dahl estimation approach. To achieve chattering-free, fast convergence, and robustness, the parameter of the SMC is adaptively tuned. Che et al. [18] considered a singularity problem with input nonlinearity by proposing observer-based adaptive integral SMC and passivity analysis. Linear matrix inequalities problems solved by using passivity conditions. Then, a singular disturbance observer is implemented, and the approximate is used in the design of adaptive law. Also, to obtain the controller parameters of integral SMC, a set of the matrix is used. Three examples are used to the verified designed controller. Erenturk [19] used two control systems such as SMC and an optimized PID controller, to control a two-mass structure. A grey estimator is applied to optimize the proposed controller. Experimental results suitably verified the applied controller performance. Jie et al. [20] proposed a novel SMC approach with terminal SMC and sliding disturbance observer for controlling a hydraulic robot manipulator. Unknown perturbations can be observed by a sliding disturbance. Furthermore, by using the hybrid controller, some time delay will have occurred. To converge the tracking error to zero, a terminal SMS is applied, which has faster speed in comparison with conventional SMC. The mentioned works mostly considered the four important phenomena by using SMC or compound control methods: high tracking performance, robustness, convergence to zero in finite time, and free-chattering.

In this work, to control of a robotic manipulator, two novel controllers are proposed. First, it is well established that the essential and the most crucial part of SMC design is how to select SMS, which is provided to response desired control performance. The trajectories must lie on the SMC. Therefore, a new SMS is proposed, which has high tracking performance and robust against external perturbations. Chattering is the main problem of the designed controller. A new reaching control law is used in the controller, which significantly will alleviate the chattering. Finally, experimental results validate that the proposed controller will alleviate chattering, which created by proposed SMC.

The rest of this work is organized as follows. The dynamic model of a robot manipulator is introduced in Section 2. Section 3 describes a new SMC. The novel reaching control law is considered in Section 4. Section 5 includes the experimental results in detail. Conclusions are provided in Section 6.

**2. A dynamic model of the robot manipulator**

Robotics manipulator is widely applicable in different fields such as industrial robots, biorobotics, and aerospace robots. Upper-limb exoskeleton robot is the one type of robot that constructed with manipulators in different positions. Therefore, if the manipulators control appropriately, they will perform suitably as a robot to assist the people who need s some help for their rehabilitation problems [22,23]. The mechanism of the proposed two degrees of freedom (2 DOF) robot manipulator is illustrated in Fig. 1.

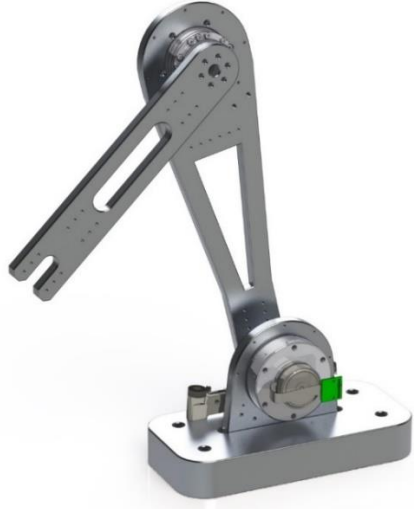

**Fig. 1.** Robot manipulator

Fig. 2 shows the schematic of the robotic manipulator. The dynamic modeling of a 2DOF robot arm is as follows [23]:

$$M(q)\ddot{q} + N(q,\dot{q})\dot{q} + G(q) = \tau \qquad (1)$$

Where $q, \dot{q}, \ddot{q} \in R^2$ illustrates the position, velocity, and acceleration of the joints, respectively. Also, in the dynamic model of the 2DOF robot manipulator, $M(q) \in R^{2\times 2}$ represented as the inertia matrix, $N(q,\dot{q}) \in R^{2\times 2}$ known as the vector of centrifugal and Coriolis forces, $G(q) \in R^{2\times 1}$ is a gravitational vector, and $\tau \in R^{2\times 1}$ the joint torques. $M(q), N(q,\dot{q})$, $G(q)$ and are provided in Appendix A.

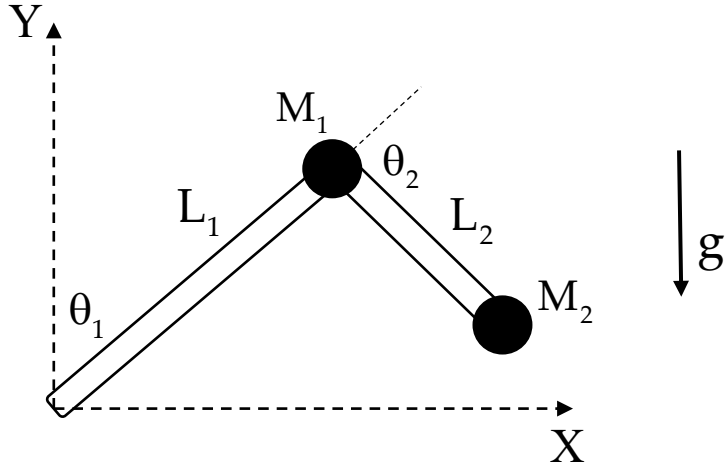

**Fig. 2.** Structure of robot manipulator.

The Eq. (1) can be expressed as:

$$\ddot{q} = M^{-1}(q)(\tau - N(q,\dot{q})\dot{q} - G(q)) = M^{-1}(q)\tau - M^{-1}(q)N(q,\dot{q})\dot{q} - M^{-1}(q)G(q) \tag{2}$$

The Eq. (2) can be written as:

$$\ddot{q} = Wu - V\dot{q} - HG(q) \tag{3}$$

Where $W = M^{-1}(q)$, $V = M^{-1}(q)N(q,\dot{q})$, and $U(t) = W\tau$. The Eq. (3) can be shown as:

$$\ddot{q} = U(t) - (V + \Delta V)\dot{q} - (W + \Delta W)G(q) + E(t) \tag{4}$$

Where $\Delta W$ and $\Delta V$ represent some uncertainties of parameter variations, $E(t)$ and describes as the external disturbances. The Eq. (4) can be demonstrated as:

$$\ddot{q} = U(t) - V\dot{q} - WG(q) + d(t) \tag{5}$$

Where $d(t) = -\Delta V\dot{q} - \Delta WG(q) + E(t)$.

## 3. Novel sliding mode control

It is well established that the essential and the most crucial part of SMC design is how to select SMS, which is provided to respond to desired control performance.

The trajectories must lie on the SMC. Therefore, a new SMS is proposed, which has high tracking performance and robust against external perturbations as:

$$f(t) = \dot{e}_r(t) + \int_0^t (k_1 sig(e_r(t)) + k_2 sig(\dot{e}_r(t))) \, d\tau \tag{6}$$

where

$$sig(e_r(t)) = |e_r(t)| sign(e_r(t))$$
$$sig(\dot{e}_r(t)) = |\dot{e}_r(t)| sign(\dot{e}_r(t))$$

Where $e_r(t) = q_d - q$ is the tracking error and $\dot{e}_r(t) = \dot{q}_d - \dot{q}$. The SMC includes two crucial cases: equivalent control and reaching control law.

To obtain the equivalent controller, the SMS should be enforced to zero ($\dot{f}(t) = 0$) as:

$$\dot{f}(t) = \ddot{e}_r(t) + k_1 sig(e_r(t)) + k_2 sig(\dot{e}_r(t)) = 0 \tag{7}$$

Substitute $\ddot{e}_r(t) = \ddot{q}_d - \ddot{q}$ in Eq. (7) generates

$$\ddot{q}_d - \ddot{q} + k_1 sig(e_r(t)) + k_2 sig(\dot{e}_r(t)) = 0 \tag{8}$$

Substitute Eq. (5) into Eq. (8) produces

$$\ddot{q}_d - U(t) + V\dot{q} + WG(q) - d(t) + k_1 sig(e_r(t)) + k_2 sig(\dot{e}_r(t)) = 0 \tag{9}$$

The equivalent control will be defined as:

$$u_{eq}(t) = \ddot{q}_d + V\dot{q} + WG(q) - d(t) + k_1 sig(e_r(t)) + k_2 sig(\dot{e}_r(t)) \tag{10}$$

When external perturbations apply to the system, the equivalent control is enabled to suppress those noises. Thus, a second control law should be defined to be robust against external perturbations. The conventional reaching control law, which has been used in several types of research [24, 25], will be selected according to Eq. (11). The reasons why reaching control implemented in most of the cases is its robustness and high tracking performance.

$$u_r(t) = K_r sign(e_r(t)) \tag{11}$$

Where $K_r$ is the positive constant. The proposed control input shows as:

$$u_{NSMC}(t) = u_{eq}(t) + u_r(t) \tag{12}$$

The Lyapunov theory is a strong tool for proving the stability of the proposed controller as:

$$L_y = \frac{1}{2} f^T(t) f(t) \tag{13}$$

Take derivative from Eq. (13) generates

$$\dot{L}_y = f^T(t) \dot{f}(t) < 0, \quad f(t) \neq 0. \tag{14}$$

When the Eq. (14) satisfy, the control system will be stable. Substitute Eq. (7) into Eq. (14) produces:

$$\dot{L}_y = f^T(t)(\ddot{e}_r(t) + k_1 sig(e_r(t)) + k_2 sig(\dot{e}_r(t))) \tag{15}$$

The Eq. (15) will be denoted as:

$$\dot{L}_y = f^T(t)(\ddot{q}_d - \ddot{q} + k_1 sig(e_r(t)) + k_2 sig(\dot{e}_r(t))) \tag{16}$$

Substitute Eq. (5) into Eq. (16) demonstrates

$$\dot{L}_y = f^T(t)(\ddot{q}_d - U(t) + V\dot{q} + WG(q) - d(t) + k_1 sig(e_r(t)) + k_2 sig(\dot{e}_r(t))) \tag{17}$$

Substitute Eq. (12) into Eq. (17) produces

$$\dot{L}_y = f^T(t)(\ddot{q}_d - u_{eq}(t) - u_r(t) + V\dot{q} + WG(q) - d(t) + k_1 sig(e_r(t)) + k_2 sig(\dot{e}_r(t))) \tag{18}$$

Substitute Eq. (10) and Eq. (11) into Eq. (18) denotes

$$\dot{L}_y = f^T(t)(\ddot{q}_d - \ddot{q}_d - V\dot{q} - WG(q) + d(t) - k_1 sig(e_r(t)) - k_2 sig(\dot{e}_r(t)) - K_r sign(e_r(t)) \\ + V\dot{q} + WG(q) - d(t) + k_1 sig(e_r(t)) + k_2 sig(\dot{e}_r(t))) \tag{19}$$

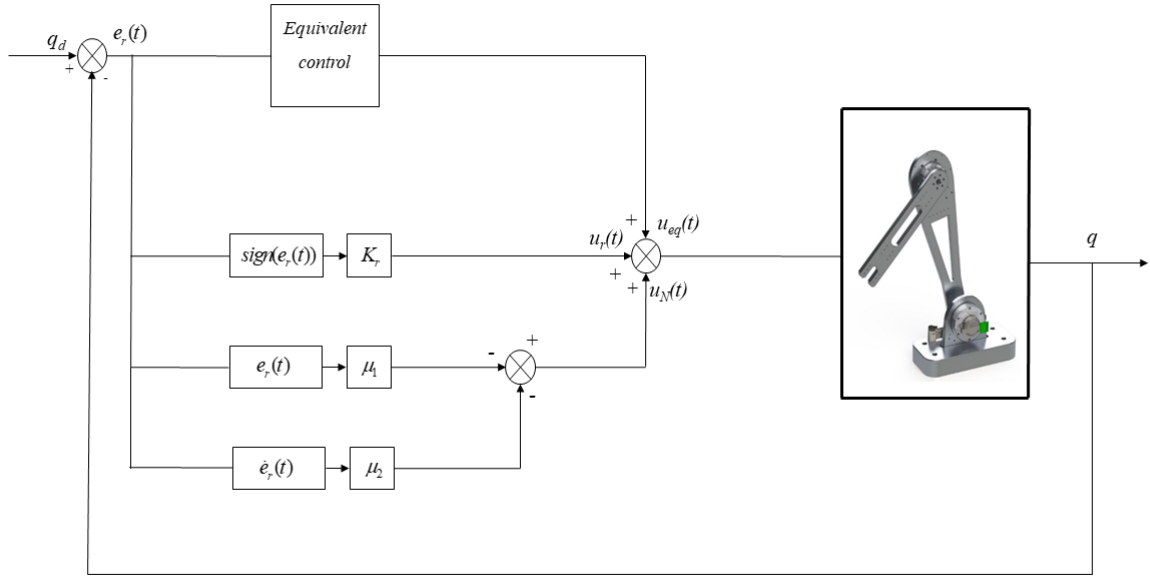

**Fig. 3**. Block diagram of the proposed controller.

Simplify Eq. (19) produces

$$\dot{L}_y = f^T(t)(-K_r sign(e_r(t))) \leq -\left|f^T(t)\right|(K_r sign(e_r(t))) \tag{20}$$

The Eq. (20) satisfies the condition of Eq. (14). Thus, the proposed control approach is stable.

## 4. New compound control law

Reaching control law is one of the important parts of designing SMC. If it selects suitably, an efficient controller will be designed. The chattering will be alleviated when the perturbation is high by choosing a suitable reaching control function. The provided reaching control law (Eq. (11)) is unable to force the system to track SMS. A new compound control function is introduced to solve the chattering-free problem. The proposed controller block diagram is illustrated in Fig. 3. The new compound control law defines as:

$$u_{NCSMC}(t) = u_{NSMC}(t) + u_N(t) \tag{21}$$

where $u_N(t)$ can be defined as:

$$u_N(t) = -\mu_1 e_r(t) - \mu_2 \dot{e}_r(t) \tag{22}$$

Where $\mu_1$ and $\mu_2$ are positive constants. The stability of the controller by using the new reaching control will be defined by substituting Eq. (21) into Eq. (17) generates:

$$\dot{L}_y = f^T(t)(\ddot{q}_d - u_{NSMC}(t) - u_N(t) + V\dot{q} + WG(q) - d(t) + k_1 sig(e_r(t)) + k_2 sig(\dot{e}_r(t))) \tag{23}$$

Also, substitute Eq. (12), Eq. (10), and Eq. (11) into Eq. (23) produces

$$\begin{aligned}\dot{L}_y &= f^T(t)(\ddot{q}_d - u_{NSMC}(t) - u_N(t) + V\dot{q} + WG(q) - d(t) + k_1 sig(e_r(t)) + k_2 sig(\dot{e}_r(t))) \\ &= f^T(t)(\ddot{q}_d - u_{eq}(t) - u_r(t) - u_N(t) + V\dot{q} + WG(q) - d(t) + k_1 sig(e_r(t)) + k_2 sig(\dot{e}_r(t))) \\ &= f^T(t)(\ddot{q}_d - \ddot{q}_d - V\dot{q} - WG(q) + d(t) - k_1 sig(e_r(t)) - k_2 sig(\dot{e}_r(t)) - K_r sign(e_r(t)) \\ &\quad + \mu_1 e_r(t) + \mu_2 \dot{e}_r(t) + V\dot{q} + WG(q) - d(t) + k_1 sig(e_r(t)) + k_2 sig(\dot{e}_r(t)))\end{aligned} \tag{24}$$

Simplify Eq. (23) generates

$$\dot{L}_y = f^T(t)(-K_r sign(e_r(t)) + \mu_1 e_r(t) + \mu_2 \dot{e}_r(t)) \leq -\left|f^T(t)\right|(K_r sign(e_r(t)) + \mu_1 |e_r(t)| + \mu_2 |\dot{e}_r(t)|) \tag{25}$$

The Eq. (25) shows that the proposed controller by using the new reaching law is stable.

## 5. Experimental results

### *5.1. Characterization of the robot manipulator*

The robot is a two degree of freedom (2 DoF) robot manipulator (Fig. 1). This robot can be widely used in different structures, such as industry and medicine. One of the most important applications of this robot can be used for rehabilitation issues when people are difficulty move their upper-limb. It can use for simulating shoulder joint vertical flexion/extension and elbow flexion/extension movements. Joint one, which includes angle 1, simulates shoulder joint vertical flexion/extension, and joint two simulates elbow flexion/extension movement.

## 5.2. Real-time setup

The experimental schematic design for the 2 DoF robot manipulator is demonstrated in Fig. 4.

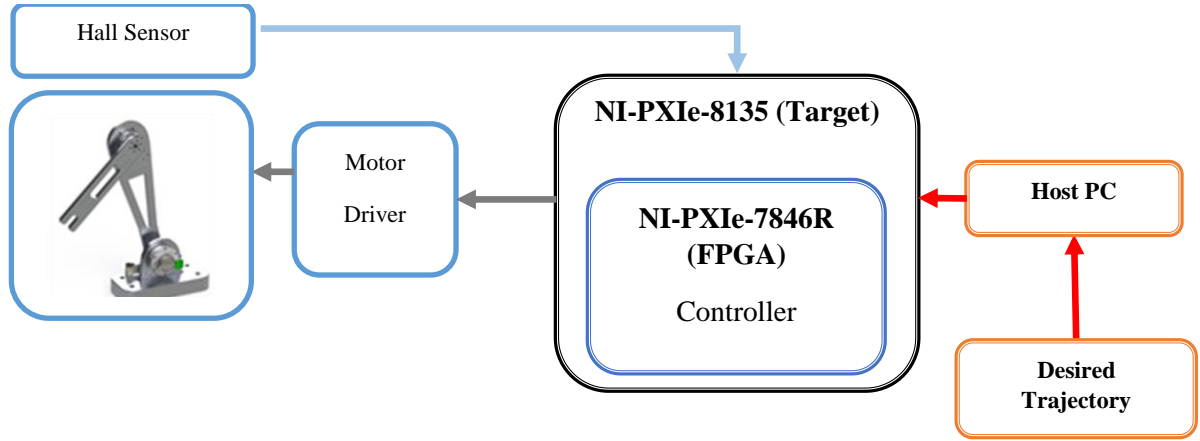

**Fig. 4.** Experimental setup

To actuate the 2 DoF robot arm, two Maxon motors (EC45) integrated with harmonic derive are attached to the joint one and joint two. In order to provide the RPM/ angle of rotation, the motors implement built-in Hall sensors. The sample that Hall sensor use is at 1ms. In order to remove high frequency, a second-order filter system is implemented (filtered parameters: $\zeta = 0.9$, and $\omega_0 = 30$ rad/s). Fig. 5 illustrates the control structure for robot arm, the setup to read the analog signals and implement the desired control via voltage/current pulses its archived by using a host PC running the latest NI LabView® connected through ethernet to a NI (National Instrument) target running a Linux-based-real-time OS that allows to process and store information at faster rates and an FPGA module as the main component with a higher sampling rate to read the encoders (Hall Sensors) from the motors and send the control current to the motor drivers. As presented in the controller development section and also in Fig. 5, the controller outputs are the robot joints' torques, which are later converted into the current commands for motor drivers. The proposed controller runs in NI-PXIe (Fig. 5, the sampling rate of 1.25 ms). As also seen in Fig. 5, a low-level Proportional Integral (PI) controller to control the desired current runs at 50μs (Fig. 5) inside the FPGA.

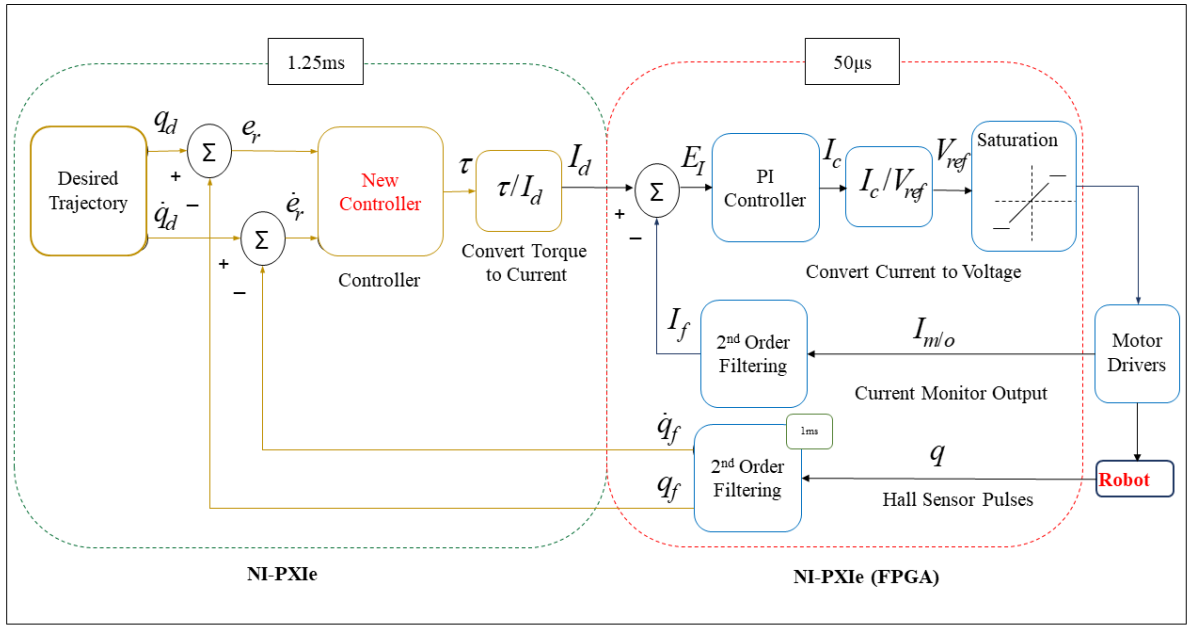

**Fig. 5**. Control architecture

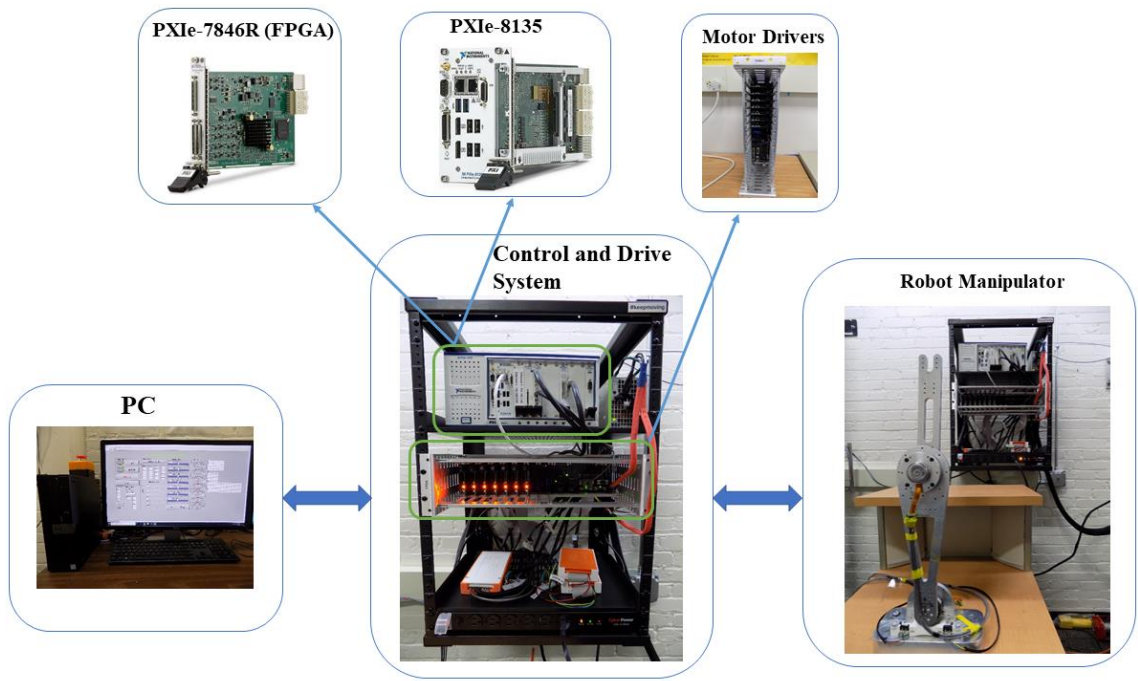

**Fig. 6**. Experimental schematic of 2 DoF robot

The feedback current signals measured from the motor drivers (at a sampling rate a1ms) are also filtered with a second-order filter (sampling parameters: ζ=0.90, and ω0=3000 rad/s). prior to being sent to the PI controller. Fig. 6 shows the experimental schematic of the 2 DoF robot.

*5.3. Results*

The robot structure properties are chosen as $L_1$=320mm, $L_2$=360mm, $m_1$=386 gr, and $m_2$=722 gr. The controller parameters are selected as $k_1 = diag\{580, 580\}$, $k_2 = diag\{50, 50\}$, $K_r = diag\{30, 30\}$, $\mu_1 = diag\{40, 40\}$, and $\mu_2 = diag\{40, 40\}$. When the robot encountered with chattering, it's not able to track the desired trajectory suitably. Also, chattering makes robot unstable. This paper proposed a new controller that significantly reduced the chattering phenomenon. Fig. 7 shows the response tracking of robot joints under the conventional sliding mode control (SMC), new SMC (NSMC) and new compound control SMC (NCSMC). The chattering phenomenon will damage the motors, which needs to be reduced by proposing some new algorithms. According to Fig. 7, the NSMC will track the desired trajectory suitably by using Cartesian trajectory tracking, but the main problems of that control are creating oscillation. By using the new reaching controller, the chattering considerably reduced. This issue proved by magnifying a part of the figure for both joints 1 and 2. Fig. 8 illustrates the position tracking error of joints 1 and 2 under SMC, NSMC and NCSMC. Fig. 9 shows the control input signals under SMC, NSMC and NCSMC. Fig. 9 demonstrates that the chattering phenomenon reduced considerably. For joint 1, for instance, the chattering reduced from 40 (N.m) to zero in t=10 (sec) approximately.

## 6. Conclusion

Designing a new control approach for reducing the chattering was the goal of this work. Chattering may damage to the motors if continuously continue when the robot is moving. Firs, a new SMC is introduced to improve the tracking performance of robot arm angles. It had a suitable tracking performance, but the huge oscillations were the main drawbacks of that controller. A new reaching control law proposed to reduce chattering created by the proposed SMC. The new reaching control algorithm constantly evaluated error values and send the correction values to the system. As a result of this, oscillation reduced considerably. The experimental results clearly verified that the new reaching control algorithm significantly reduced the chattering phenomenon.

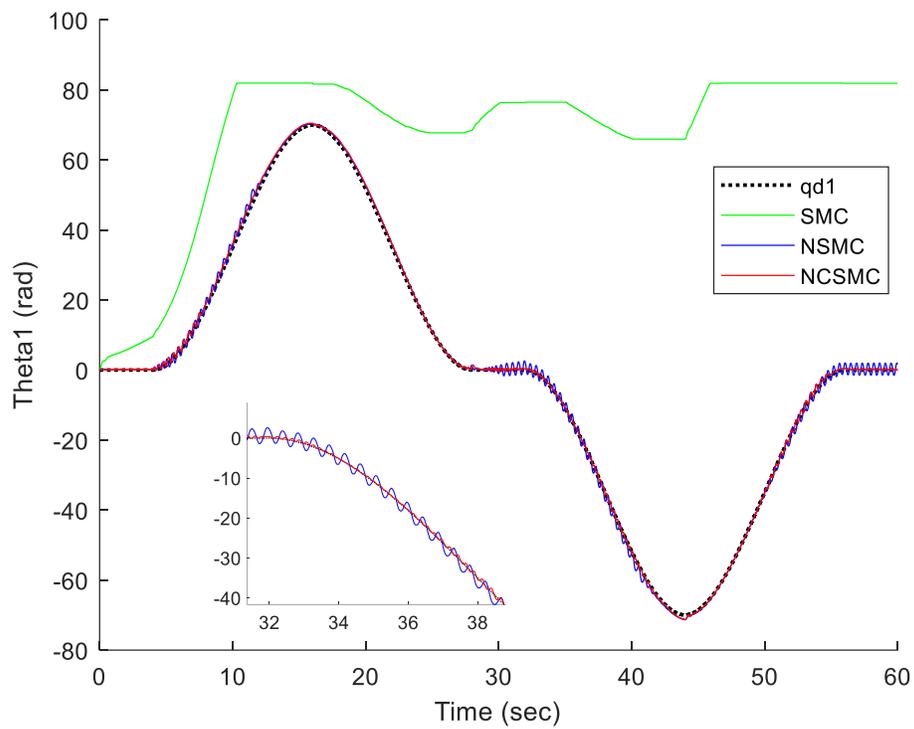

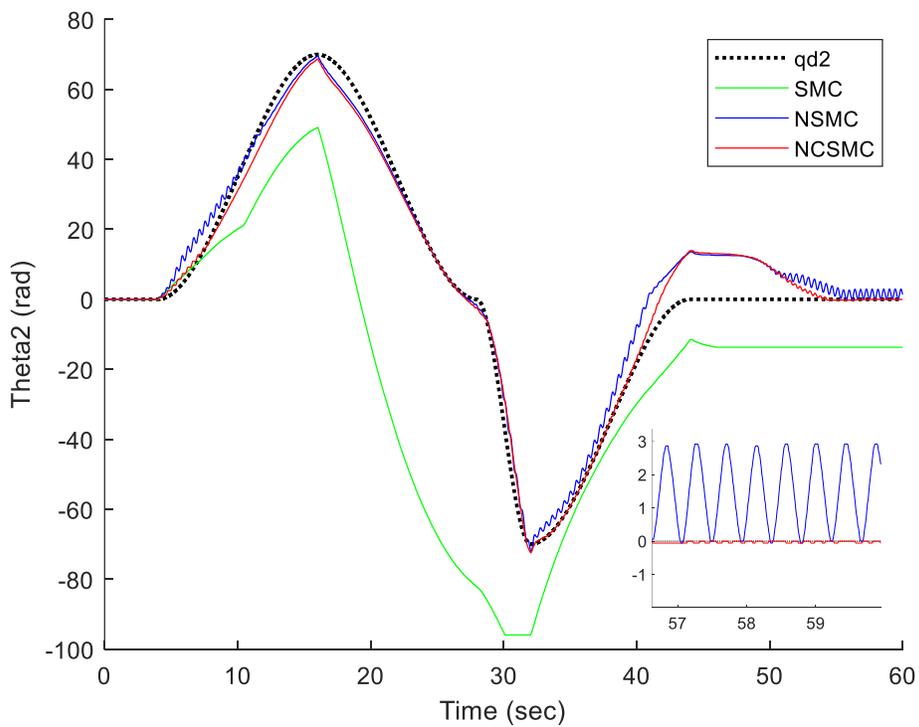

**Fig. 7**. Tracking of joints under SMC, NSMC and NCSMC.

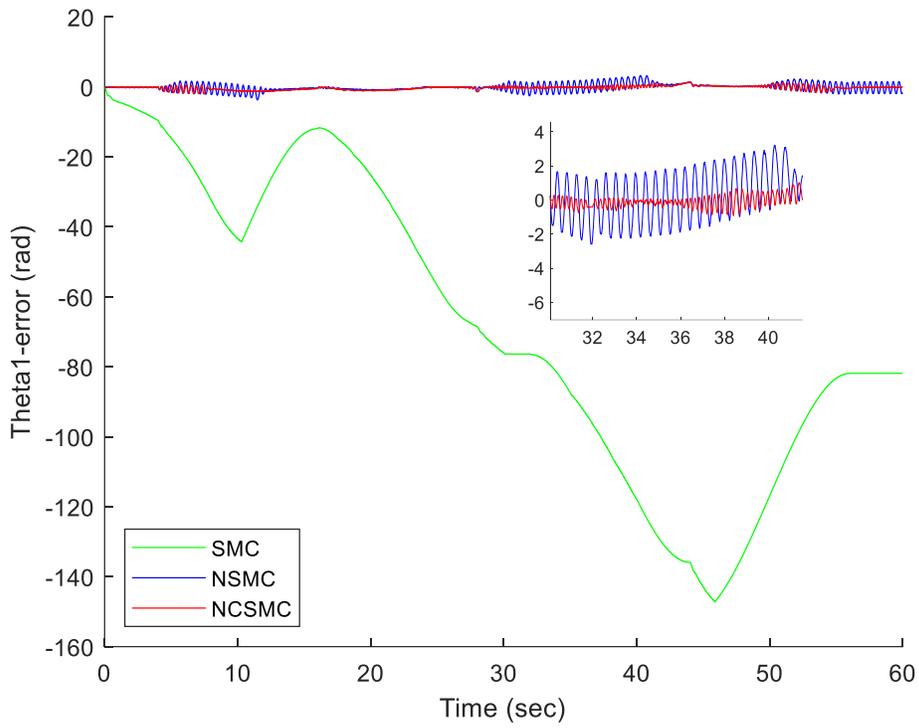

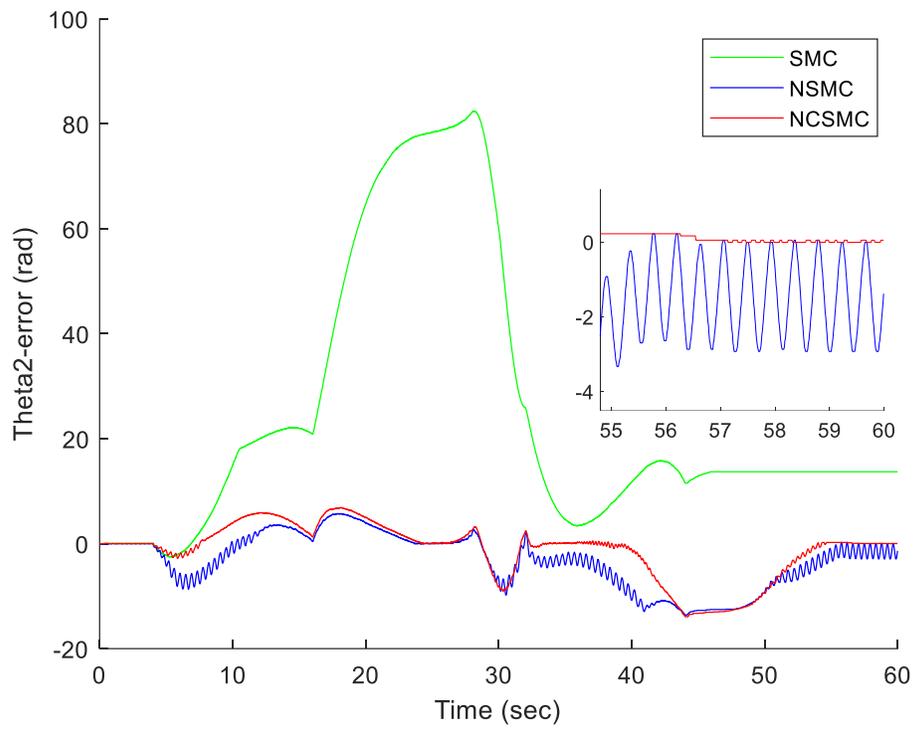

**Fig. 8**. Tracking error of joints under SMC, NSMC and NCSMC.

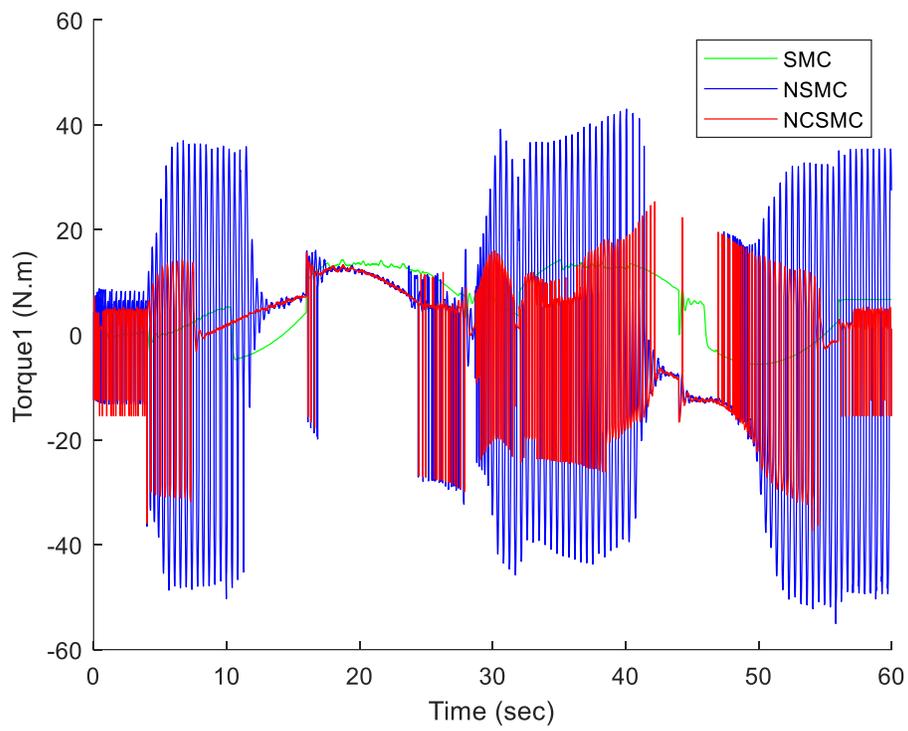

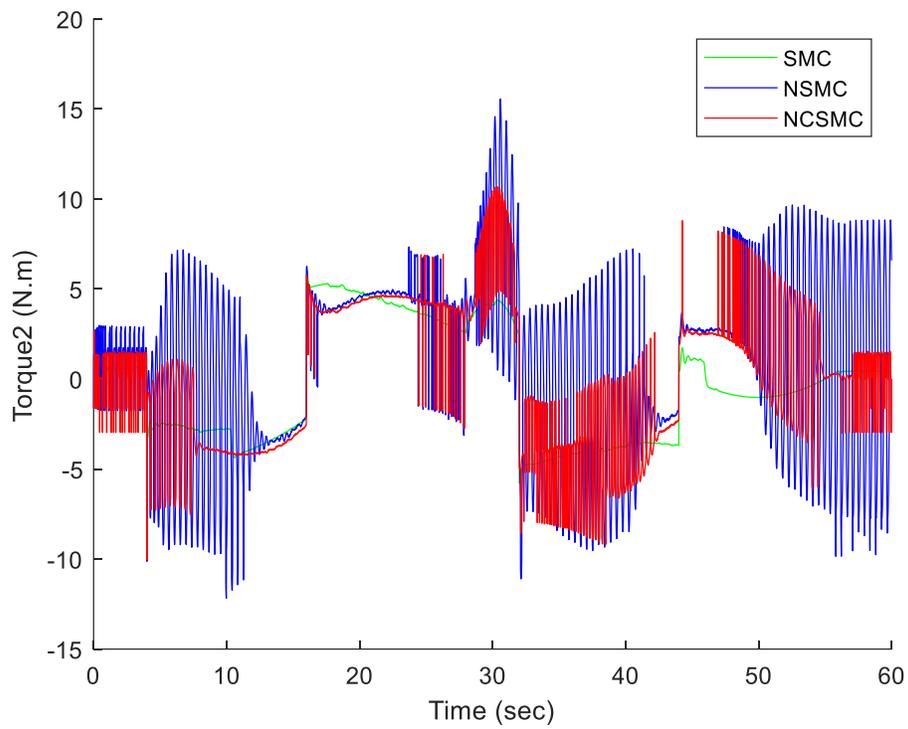

**Fig. 9**. Control input signals under SMC, NSMC and NCSMC.

**Appendix A**

$$q = \begin{bmatrix} \theta_1 \\ \theta_2 \end{bmatrix}$$

$$M(q) = \begin{bmatrix} (M_1 + M_2)L_1^2 + M_2 L_2^2 + 2M_2 L_1 L_2 \cos\theta_2 & M_2 L_2^2 + M_2 L_1 L_2 \cos\theta_2 \\ M_2 L_2^2 + M_2 L_1 L_2 \cos\theta_2 & M_2 L_2^2 \end{bmatrix}$$

$$N(q, \dot{q}) = \begin{bmatrix} -M_2 L_1 L_2 \sin\theta_2 (2\dot{\theta}_1 \dot{\theta}_2 + \dot{\theta}_2^2) \\ -M_2 L_1 L_2 \sin\theta_2 \dot{\theta}_1 \dot{\theta}_2 \end{bmatrix}$$

$$G(q) = \begin{bmatrix} -(M_1 + M_2)gL_1 \sin\theta_1 - M_2 g L_2 \sin(\theta_1 + \theta_2) \\ -M_2 g L_2 \sin(\theta_1 + \theta_2) \end{bmatrix}$$

$$\tau = \begin{bmatrix} \tau_1 \\ \tau_2 \end{bmatrix}$$